# A Quantized Representation of Probability in the Brain

James Tee, *Member, IEEE*, and Desmond P. Taylor, *Life Fellow, IEEE*

*Abstract*—Conventional and current wisdom assumes that the brain represents probability as a continuous number to many decimal places. This assumption seems implausible given finite and scarce resources in the brain. Quantization is an information encoding process whereby a continuous quantity is systematically divided into a finite number of possible categories. Rounding is a simple example of quantization. We apply this information theoretic concept to develop a novel quantized (i.e., discrete) probability distortion function. We develop three conjunction probability gambling tasks to look for evidence of quantized probability representations in the brain. We hypothesize that certain ranges of probability will be lumped together in the same indifferent category if a quantized representation exists. For example, two distinct probabilities such as 0.57 and 0.585 may be treated indifferently. Our extensive data analysis has found strong evidence to support such a quantized representation: 59/76 participants (i.e., 78%) demonstrated a best fit to 4-bit quantized models instead of continuous models. This observation is the major development and novelty of the present work. The brain is very likely to be employing a quantized representation of probability. This discovery demonstrates a major precision limitation of the brain's representational and decision-making ability.

## I. INTRODUCTION

Humans tend to subjectively overestimate small probabilities (e.g., dying in an airplane crash, winning the Powerball jackpot lottery) and underestimate large probabilities (e.g., getting lung cancer from smoking cigarettes) when making decisions under risk and uncertainty. These subjective probabilities are typically modeled over the range [0,1] using a probability distortion (or weighting) function [4]-[7]. It is commonly assumed that subjective probability is represented in the brain in continuous form (i.e., a Real number that can take any value between 0 and 1). This assumption may result in inaccurate analytical findings if decisions are made using a non-continuous (i.e., finite precision) representation. For example, if an individual were to choose between a $100 lottery of a winning probability of 0.89721, versus a $150 lottery of a winning probability of 0.50219, s/he will typically round the numbers when making the choice [8]. The choice is then greatly simplified to a $100 lottery with a winning probability of 0.9, versus a $150 lottery with a winning probability of 0.5. Analysis performed using a continuous representation is, strictly speaking, inaccurate because 0.89721 is actually treated indifferently to 0.9. Rounding implies that a range of probabilities is treated as being the same (i.e., indifferent). This analytical inconvenience is often ignored because standard (i.e., continuous) models cannot easily take into account such indifference. While Kahneman and Tversky [8] expected rounding (i.e., "editing"), they did not quantify it.

Quantization [9] is the term describing an encoding process in which a continuous quantity (i.e., Real number or analog signal) is systematically divided into a finite number of possible categories. We note that the term "quantization" is loosely synonymous to terminologies such as discretization, chunking, categorization and classification. The term "discrete" is common in neuroscience, whereas the term "quantized" is more common in engineering. Here, we use both terms interchangeably. Rounding a number is the oldest example of quantization [10]. Oliver, Pierce and Shannon [11] employed quantization to convert continuous signals (e.g., audio/voice) to their digital (i.e., binary) forms to enable the maximum possible efficiency of data compression and communications [12], thus laying the theoretical foundations for almost all modern communications systems. For binary quantization, the total number of categories equals to $2^n$, where $n$ is the quantization precision expressed in bits. For example, suppose that we are encoding probabilities (i.e., Real numbers in the interval [0,1]). If $n$ is one, then we have only two (i.e., $2^1$) output categories. We might encode all quantities in the interval [0, 0.5] as the first output category, with the remaining probabilities (0.5, 1] forming the second category. The distinction between substantially different probabilities such as 0.001 and 0.5 is lost in the encoding (i.e., 0.001 and 0.5 are mapped indifferently to the same category). When $n$ is large, such as 24 bits, we can encode each observation to $2^{24}$ possible very finely spaced categories, resulting in only a small loss of precision compared to the underlying Real

Manuscript accepted on October 18, 2019. This work was supported by grant EY019889 from the National Institutes of Health. This paper was presented in part at the 2013 Annual Meeting of the Society for Neuroeconomics in Lausanne, Switzerland [1]. Parts of this paper were first published in James Tee's PhD dissertation research, undertaken and successfully completed at the Department of Psychology of New York University [2]. An earlier version of this paper was also published as an arXiv preprint [3].

James Tee was previously with the Department of Psychology, New York University. He is now with the Communications Research Group, Department of Electrical & Computer Engineering, University of Canterbury, Private Bag 4800, Christchurch 8020, New Zealand (email: james.tee@canterbury.ac.nz).

Desmond P. Taylor is with the Communications Research Group, Department of Electrical & Computer Engineering, University of Canterbury, Private Bag 4800, Christchurch 8020, New Zealand (email: desmond.taylor@canterbury.ac.nz).

number. Thus, a larger number of bits represents a higher level of precision (in the engineering sense). In fact, a continuous quantity is the special case of $n = \infty$. Note that, in a typical electrophysiology experimental data recording equipment, continuous neural signals (i.e., electrical voltage data in neurons) from electrodes are passed on to a high precision (e.g., 16-bit) analog-to-digital converter (ADC, also known as a quantizer) before being stored as a digital/discrete data file [13].

In this paper, we conduct three human behavioral experiments and analyze the experimental results based on a novel quantized (i.e., discrete) probability distortion function, in comparison to a conventional (i.e., continuous) probability distortion function. Our quantization approach is, in part, motivated by Miller [14], who hypothesized that the magical number 7, plus or minus 2, represents the brain's capacity limit for processing information. He reported experimental evidence showing that the brain's inherent accuracy limit for absolute judgments of the position of a pointer in a linear interval is about 3.25 bits, suggesting that the brain categorizes (or quantizes) stimuli into 9 or 10 ($2^{3.25} = 9.5$) discrete categories (instead of a continuous or infinite number of categories). We hypothesize that the brain has a precision limit in representing probabilities. This suggests that subjective probability estimates in humans may not be represented as a continuum between 0 and 1, but rather, as a discrete set of categories (e.g., very low, low, medium, high, and very high), such that certain ranges of distinct probability values (e.g., from 0.1 to 0.3) will be treated indifferently (i.e., quantized into the same category, such as "low"). Our approach is related to the quantized model of perception proposed by Sun, Wang, Goyal and Varshney [15], and the recent work on quantized prior probabilities applied to Bayesian models of human decision-making by Varshney and Varshney [16]. The work here also builds on Tee and Taylor [17] where the neurobiological basis for discrete representation of information in the brain was established using communications theory. Since quantization is the fundamental basis for most communications systems operating over a noisy channel, this notion is directly relevant and applicable to neurobiological communications in the brain.

## II. A NOVEL QUANTIZED (DISCRETE) MODEL OF PROBABILITY

At present, subjective probability is modeled by several different probability distortion (or weighting) functions [4]-[7], all of which assume continuous representation. We will base the present work on the two-parameter continuous probability function due to Prelec [6], defined as:

$$w(x) = e^{-\delta(-\ln x)^\gamma}$$

where $x$ is the objective probability, $w(x)$ is the subjective probability, $\delta > 0$ and $\gamma > 0$. Depending on the values of $\delta$ and $\gamma$, this function can be either S- or inverse S-shaped. Fig. 1C shows an example of a continuous Prelec function.

We quantize [9] the continuous Prelec function, resulting in a novel quantized (discrete) version:

$$Q_n[w(x)] = Q_n[e^{-\delta(-\ln x)^\gamma}]$$

where $w(x)$ is a continuous Prelec function, $n$ is the number of bits and $Q_n[.]$ is a quantization function. The quantization function divides the continuous probability space $[0,1]$ into $2^n$ equally spaced steps, with each step size being equal to $2^{-n}$. The step sizes of the corresponding values of $n$ are shown in Table I, third column. Using $y = x$ as a simple example, $y = Q_3[x]$ divides y into $2^3 = 8$ equally spaced steps. These are plotted in Fig. 1A and Fig. 1B. The quantization (or systematic rounding) effect is further demonstrated in the 3-bit graph (Fig. 1B), where it is shown that two distinct values of $x$ (i.e., 0.39 in blue, and 0.47 in green) map to the same (indifferent) value of $y$ ($\approx 0.42$ in purple). In the same manner, a continuous Prelec function, $w_a(x)$, can be quantized to produce $y = Q_n[w_a(x)]$. This quantized function is specified by 3 parameters: $\delta$, $\gamma$ and $n$. Fig. 1C and Fig. 1D show a continuous Prelec function ($\delta = 0.9$, $\gamma = 0.6$) and its corresponding 3-bit quantized version respectively. We note that $n$ determines the step size, while $\delta$ and $\gamma$ determine where these steps occur. If $n$ is large enough, the quantized Prelec function becomes almost indistinguishable from the continuous version. A continuous Prelec function is a special case of a quantized Prelec function with $n = \infty$.

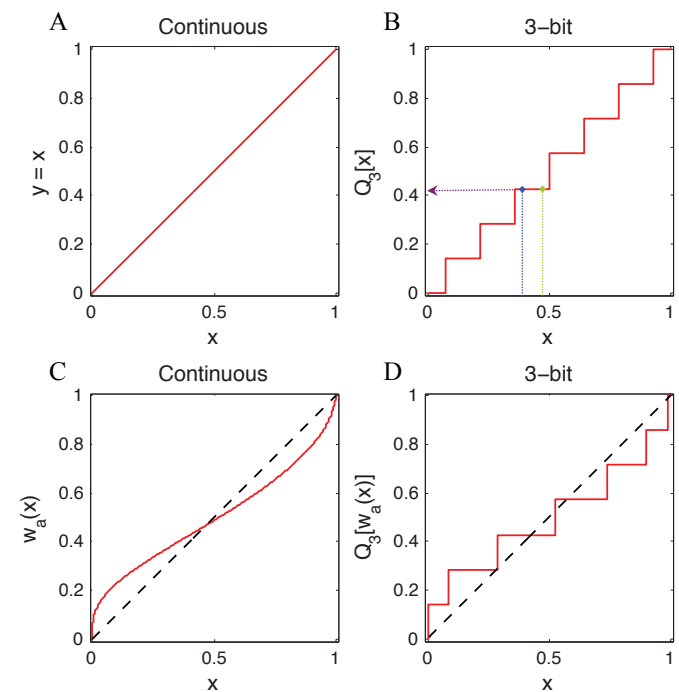

Fig. 1. Examples of continuous functions and their corresponding quantized versions. (A) Continuous linear function $y = x$. Note that this is mathematically equivalent to the case of a Prelec function with $\delta = 1$ and $\gamma = 1$. (B) 3-bit quantized version of linear function, $Q_3[x]$. The arrows show the quantization effect, where two different values of $x$ (i.e., 0.39 in blue, and 0.47 in green) both map to the same (indifferent) value of $y$ (i.e., $\approx 0.42$ in purple). (C) Continuous Prelec function, $w_a(x)$, with $\delta = 0.9$ and $\gamma = 0.6$. (D) 3-bit quantized version of the Prelec function, $Q_3[w_a(x)]$.

TABLE I
THE SIZE OF EACH STEP CORRESPONDING TO THE QUANTIZATION PRECISION

| Precision (bit) | Number of steps (categories) | Size of each step (category) | Angle on a wheel (degrees) |
| --- | --- | --- | --- |
| 1 | $2^1 = 2$ | 0.5 | 180 |
| 2 | $2^2 = 4$ | 0.25 | 90 |
| 3 | $2^3 = 8$ | 0.125 | 45 |
| 4 | $2^4 = 16$ | 0.0625 | 22.5 |
| 5 | $2^5 = 32$ | 0.03125 | 11.25 |
| 6 | $2^6 = 64$ | 0.015625 | 5.625 |
| 7 | $2^7 = 128$ | 0.0078125 | 2.8125 |
| 8 | $2^8 = 256$ | 0.00390625 | 1.40625 |
| 9 | $2^9 = 512$ | 0.001953125 | 0.703125 |
| 10 | $2^{10} = 1024$ | 0.0009765625 | 0.3515625 |
| 20 | $2^{20} = 1048576$ | $9.53674 \times 10^{-7}$ | 0.000343323 |

## III. METHODS

### A. Overview

Fig. 2 provides an overview of the experimental setup and data fitting process. There are 3 experiments in total. Each experiment is a gambling task based on roulette wheels. In each experiment, a participant is presented with the gambling task on a touchscreen computer monitor. The participant's decision choices are recorded as experimental data. Maximum likelihood estimation is used to fit experimental data to quantized probability models. Nested hypothesis statistical tests are performed at a 0.05 level to reduce the number of model parameters. This gives us the quantized probability model that best approximates the participant's decision-making process. Details of the experimental setup and data fitting process are further described in the rest of this section.

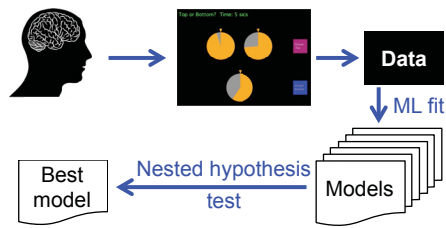

Fig. 2. Overview of the experimental setup and data fitting process.

### B. Experiment 1

This is a 2-event conjunction task, as shown in Fig. 3. During each trial of this task, a participant has up to 10 seconds to choose between a single roulette wheel with probability $r$ of success and a pair of independent roulette wheels with probabilities $x$ and $z$ of success. The participant's estimates of $x$, $z$, and $r$ were based on visual judgments of the fraction of each roulette wheel colored gold. After the participant has made his/her choice, the wheels were spun simultaneously at random speeds, and the participant won a monetary prize if he: (i) chose the single wheel and it stopped in the gold, or, (ii) chose the pair of wheels and both wheels stopped in the gold. A staircase procedure [18] [19] was used to estimate the $r$ for which the participant chose the pair as often as the single wheel: $r \sim (x, z)$. Twelve experimental conditions (i.e., $x$, $z$ pairs) were presented (see Fig. 3). By design, $x$ is always greater than or equal to $z$. We used a one-up/one-down staircase procedure [18] [19] with 50 trials per experimental condition such that experiment 1 consisted of 600 trials. Successive experimental conditions were presented in a random order. The horizontal order of the "large" (i.e., $x$), and "small" (i.e., $z$) wheels was randomized to mitigate order effects. The vertical placement (i.e., top or bottom) of the staircase wheel (i.e., $r$) and the conjunction wheels (i.e., an $x$, $z$ pair) was also randomized. Nine mandatory 30-second breaks were spaced uniformly throughout the experiment. The experiments were implemented using Psychtoolbox [20] [21], carried out on a 19" touchscreen powered by Windows XP.

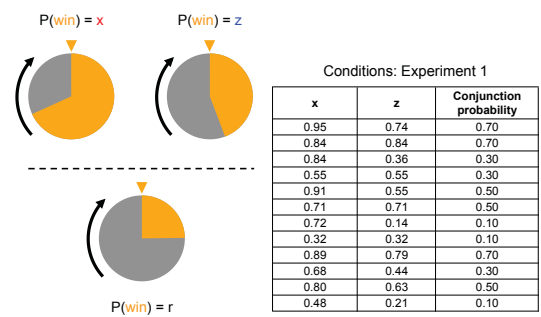

Fig. 3. The 2-event (or 2-wheel) conjunction task of experiment 1, and its 12 experimental conditions.

### C. Experiment 2

Here, we ran the same design as experiment 1, except for the experimental conditions (i.e., $x$, $z$ pairs) where we employed a different set of 15 experimental conditions (see Fig. 4). With 50 trials per experimental conditions, experiment 2 consisted of 750 trials in total. Ten mandatory 30-second breaks were spaced uniformly throughout the experiment.

Conditions: Experiment 2

| x | z | Conjunction probability |
| --- | --- | --- |
| 0.26 | 0.15 | 0.04 |
| 0.42 | 0.15 | 0.06 |
| 0.42 | 0.31 | 0.13 |
| 0.58 | 0.15 | 0.09 |
| 0.58 | 0.31 | 0.18 |
| 0.58 | 0.47 | 0.27 |
| 0.74 | 0.15 | 0.11 |
| 0.74 | 0.31 | 0.23 |
| 0.74 | 0.47 | 0.35 |
| 0.74 | 0.63 | 0.47 |
| 0.90 | 0.15 | 0.14 |
| 0.90 | 0.31 | 0.28 |
| 0.90 | 0.47 | 0.42 |
| 0.90 | 0.63 | 0.57 |
| 0.90 | 0.79 | 0.71 |

Fig. 4. The 15 experimental conditions of experiment 2.

### D. Experiment 3

The design setup for this experiment is identical to the previous experiments, except that it is a 3-event conjunction task (see Fig. 5). Twenty experimental conditions were employed (see Fig. 5). By design, $x > y > z$. With 50 trials per experimental condition, experiment 3 consisted of 1000 trials in total. The horizontal order of the "large" (i.e., $x$), "midsize" (i.e., $y$) and "small" (i.e., $z$) wheels was randomized to mitigate order effects. Thirteen mandatory 30-second breaks were spaced uniformly throughout the experiment.

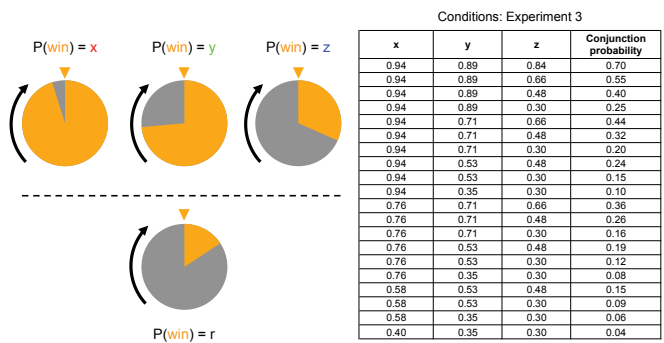

Fig. 5. The 3-event (or 3-wheel) conjunction task of experiment 3, and its 20 experimental conditions.

### E. Participants

There was a total of 87 distinct participants: 23 participants for experiment 1; 21 for experiment 2; and 44 for experiment 3; one participant participated in both experiments 1 and 3. All participants were naïve, and aged between 18 and 36 years old. There were 53 females and 34 males. Participants were paid $12 per hour. At the end of each experiment, a previously completed trial was randomly drawn, and if this trial was one that the participant had previously won, a $10 bonus was paid.

### F. Conjunction fallacy

When two events, X and Z, are independent, their conjunction (2-event) probability is the product of the constituent probabilities: $P(X \& Z) = P(X)P(Z)$. Mathematically, a conjunction probability cannot be greater than the probability of its constituents, such that $P(X \& Z) \leq P(X) \leq P(Z)$. In the "Linda the bank teller" study of Tversky and Kahneman [22], they found that 85% of participants (i.e., 75/88 participants) violated this ordinal rule, thereby committing conjunction fallacy. To investigate conjunction fallacy, we used a one-tailed binomial test against the ordinal rule at a 0.05 level of significance.

### G. Data fitting

Maximum likelihood estimation [23] was employed to fit the experimental data to each model. Successive trials in each experiment were presented in a randomized fashion, such that the preceding trial is independent from the current trial. Due to the trial-by-trial independence arising from this randomization, the prior probability becomes a constant scaling factor, and therefore, the maximum a posteriori (MAP) estimation is simplified to a maximum likelihood estimation with the constant (i.e., prior probability) being omitted (since it is merely a scaling factor). All estimations were performed in Matlab operating in double precision (Matlab's default). The key difference between fitting continuous and quantized models is that the output of the continuous models uses double precision values, whereas the output of the quantized models is restricted to a limited number of possible values (i.e., categories) as constrained by the quantization function. To simplify the data fitting process, we used integer numbers of bits for quantization. The data fitting process was verified for validity and reliability using simulations and bootstraps (see details in Section V on Data Fitting Safeguards).

### H. Nested hypothesis tests

Nested hypothesis tests [24] were conducted at a 0.05 level of significance for the purpose of parameter reduction (i.e., minimize potential over-fitting effects). For the 2-wheel task of experiments 1 and 2, we used two quantized Prelec functions to model the conjunction:

$$r = Q_k[w_a(x)]Q_n[w_c(z)]$$

where $Q_k[w_a(x)]$ and $Q_n[w_c(z)]$ are the quantized Prelec functions for the conjunction pair (for wheel $x$, $k$ is the precision and $w_a$ is the continuous Prelec function; for wheel $z$, $n$ is the precision and $w_c$ is the Prelec function). This model was applied to a hierarchical sequence of nested hypothesis tests to determine whether:
1) the participants were distorting $x$ and $z$ at all (i.e., homogeneous linear case, where estimated probabilities are objective, not subjective);
2) both $x$ and $z$ were subjectively distorted using the same Prelec function (i.e., homogeneous Prelec case);
3) $x$ and $z$ were subjective distorted using different Prelec functions (i.e., non-homogeneous Prelec case).

All nested models employed for the 2-wheel task of experiments 1 and 2 are summarized in Fig. 6.

| | | Distortion function | | |
|---|---|---|---|---|
| | | Homogeneous (i.e. same for x and z) | | Non-homogeneous (i.e. different for x and z) |
| | | Linear | Prelec | Prelec |
| Precision | Same for x and z | $Q_k[x]Q_k[z]$ | $Q_k[w_a(x)]Q_k[w_a(z)]$ | $Q_k[w_a(x)]Q_k[w_c(z)]$ |
| | Different for x and z | $Q_k[x]Q_n[z]$ | $Q_k[w_a(x)]Q_n[w_a(z)]$ | $Q_k[w_a(x)]Q_n[w_c(z)]$ |

Fig. 6. Nested hypothesis tests of models used in the 2-wheel task of experiments 1 and 2.

For the 3-wheel task of experiment 3, we used three quantized Prelec functions to model the conjunction:

$$r = Q_k[w_a(x)]Q_m[w_b(y)]Q_n[w_c(z)]$$

where $Q_k[w_a(x)]$, $Q_m[w_b(y)]$ and $Q_n[w_c(z)]$ are the corresponding quantized Prelec functions for the conjunction triplet. A similar hierarchical sequence of nested hypothesis tests was applied. All nested models employed for the 3-wheel task are summarized in Fig. 7.

| | | Distortion function | | |
|---|---|---|---|---|
| | | Homogeneous (i.e. same for x, y and z) | | Non-homogeneous (i.e. different for x, y and z) |
| | | Linear | Prelec | Prelec |
| Precision | Same for x, y and z | $Q_k[x]Q_k[y]Q_k[z]$ | $Q_k[w_a(x)]Q_k[w_a(y)]Q_k[w_a(z)]$ | $Q_k[w_a(x)]Q_k[w_b(y)]Q_k[w_c(z)]$ |
| | Different for x, y and z | $Q_k[x]Q_m[y]Q_n[z]$ | $Q_k[w_a(x)]Q_m[w_a(y)]Q_n[w_a(z)]$ | $Q_k[w_a(x)]Q_m[w_b(y)]Q_n[w_c(z)]$ |
| | 4 bits for each of x, y and z | $Q_4[x]Q_4[y]Q_4[z]$ | $Q_4[w_a(x)]Q_4[w_a(y)]Q_4[w_a(z)]$ | $Q_4[w_a(x)]Q_4[w_b(y)]Q_4[w_c(z)]$ |

Fig. 7. Nested hypothesis tests of models used in the 3-wheel task of experiment 3.

## I. Ethical approval

All human experimental procedures were performed at New York University in accordance with the guidelines of the National Institutes of Health and approved by the University Committee on Activities Involving Human Subjects (UCAIHS) serving as New York University's Institutional Review Board (IRB). The experimental protocol was explained to all participants, after which they provided their written informed consent.

## IV. RESULTS

### A. Experiment 1

Three out of 23 participants repeatedly committed conjunction fallacy. For the rest of the paper, all conjunction fallacy participants are excluded. Data fitting is applied to the remaining 20 participants. Fig. 8 shows the data fit for a single sample homogeneous participant. The quantized model is fitted from 1 to 64 bits, and the minimum point of the negative log likelihood was found to occur at 4 bits, indicating the point of best fit. [We note here that the minimum point of the negative log likelihood is equivalent to the maximum point of the log likelihood (i.e., maximum likelihood estimation process). The negative log likelihood is used here as the objective function, instead of log likelihood, in order to utilize Matlab's *fminsearch* optimization function, in the absence of a convenient "*fmaxsearch*" function.

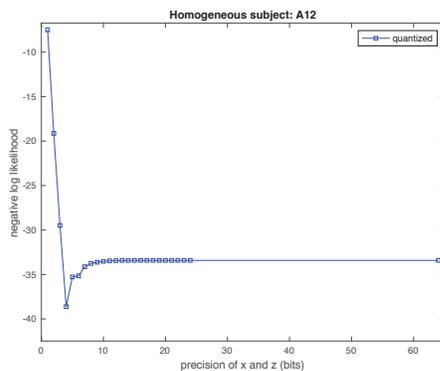

Fig. 8. Negative log likelihood of model fits for one sample participant for the 2-wheel task of experiment 1.

Fig. 9A plots a heatmap of the negative log likelihood of the maximum likelihood estimation process for another sample participant. Precisions range from 1 to 10 bits, with darker shades showing inferior fits and lighter shades superior fits. The highest precision is located at the bottom right corner where both $x$ and $z$ are 10 bits. We note that, perceptually, 10 bits of precision has very fine step sizes of $2^{-10} \approx 0.001$ (see Table I). This corresponds to about 0.35 degrees on the wheel (see Table I), which is almost indiscernible to the naked eye. As the precision decreases (i.e., moving from the bottom right corner toward the middle), the fit improves. The best fit occurs where $x$ and $z$ are both 5 bits (i.e., white square). This trend provided the first clue that a quantized model may be a better fit than a continuous model.

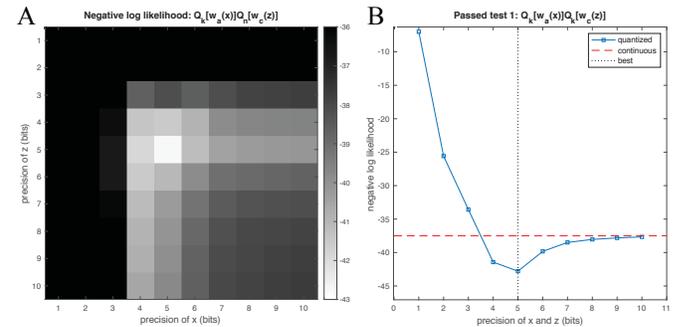

Fig. 9. Negative log likelihood of model fits for one sample participant for the 2-wheel task of experiment 1. (A) The heatmap for the 6-parameter $Q_k[w_a(x)]Q_n[w_c(z)]$ model prior to the nested hypothesis tests. (B) The graph for the 5-parameter $Q_k[w_a(x)]Q_k[w_c(z)]$ model after passing nested hypothesis test 1, showing best fit at 5 bits of precision.

In terms of nested hypothesis tests, none of the 20 participants passed the linear case (i.e., homogeneous linear case with no subjective distortions), 6 participants are homogeneous (i.e., same Prelec function subjective distortion for $x$ and $z$) and 14 participants are non-homogeneous (i.e., different Prelec functions subjective distortions for $x$ and $z$). Details are shown in Table II. Interestingly, 17 participants (who passed nested hypothesis test 1) had the same precision for $x$ and $z$ ($Q_k[.]Q_k[.]$ models). The negative log likelihood for the same sample participant is plotted in Fig. 9B. This participant passed the test for non-homogeneous Prelec distortion with $x$ and $z$ having the same precision (i.e., $Q_k[w_a(x)]Q_k[w_c(z)]$), which is why the plot is a line instead of a heatmap. The dashed horizontal red line shows the negative log likelihood value for the continuous model. As the precision decreases, the fit improves, with the best fit occurring at 5 bits. This characteristic shape of the negative log likelihood for the quantized case was found for all participants.

TABLE II
RESULTS OF NESTED HYPOTHESIS TEST 1 FOR EXPERIMENT 1

| Models | Homogeneous? (i.e. same distortion function for x and z) | Same precision for x and z? | No. of free parameters | Degrees of freedom | No. of subjects passing nested test 1 |
|---|---|---|---|---|---|
| $Q_k[x]Q_k[z]$ | Yes: linear | Yes | 1 | 5 | 0 |
| $Q_k[x]Q_n[z]$ | Yes: linear | No | 2 | 4 | 0 |
| $Q_k[w_a(x)]Q_k[w_a(z)]$ | Yes: Prelec-2 | Yes | 3 | 3 | 6 |
| $Q_k[w_a(x)]Q_n[w_a(z)]$ | Yes: Prelec-2 | No | 4 | 2 | 0 |
| $Q_k[w_a(x)]Q_k[w_c(z)]$ | No: Prelec-2 | Yes | 5 | 1 | 11 |
| $Q_k[w_a(x)]Q_n[w_c(z)]$ | No: Prelec-2 | No | 6 | - | 3 failed all tests |

There are 20 participants in total.

TABLE III
RESULTS OF NESTED HYPOTHESIS TEST 2 FOR EXPERIMENT 1

| Models | Homogeneous? (i.e. same distortion function for x and z) | Same precision for x and z? | No. of free parameters | Degrees of freedom | No. of subjects passing nested test 2 |
|---|---|---|---|---|---|
| $Q_4[x]Q_4[z]$ | Yes: linear | Yes: 4 bits | 0 | 6 | 0 |
| $Q_k[x]Q_k[z]$ | Yes: linear | Yes | 1 | 5 | 0 |
| $Q_k[x]Q_n[z]$ | Yes: linear | No | 2 | 4 | 0 |
| $Q_4[w_a(x)]Q_4[w_a(z)]$ | Yes: Prelec-2 | Yes: 4 bits | 2 | 4 | 3 |
| $Q_k[w_a(x)]Q_k[w_a(z)]$ | Yes: Prelec-2 | Yes | 3 | 3 | 3 |
| $Q_k[w_a(x)]Q_n[w_a(z)]$ | Yes: Prelec-2 | No | 4 | 2 | 0 |
| $Q_4[w_a(x)]Q_4[w_c(z)]$ | No: Prelec-2 | Yes: 4 bits | 4 | 2 | 11 |
| $Q_k[w_a(x)]Q_k[w_c(z)]$ | No: Prelec-2 | Yes | 5 | 1 | 1 |
| $Q_k[w_a(x)]Q_n[w_c(z)]$ | No: Prelec-2 | No | 6 | - | 2 failed all tests |

There are 20 participants in total. Nested hypothesis test 2 contains three additional 4-bit models.

The distributions of the best fit precisions of wheels $x$ and $z$ for all participants are plotted in Fig. 10, confirming that the quantized models fit the experimental data better than the continuous models. The best fit precisions range from 2 to 5 bits, with modes of 4 bits.

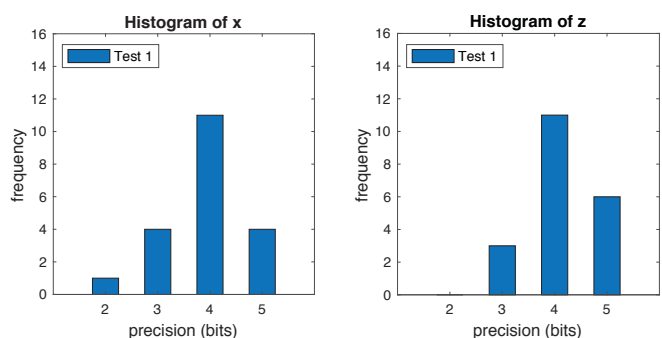

Fig. 10. Distributions of best fit precisions of wheels $x$ and $z$ for all participants from experiment 1 after passing nested hypothesis test 1. This confirms that quantized models do indeed fit the experimental data better than continuous models, with a mode of 4 bits for both wheels $x$ and $z$.

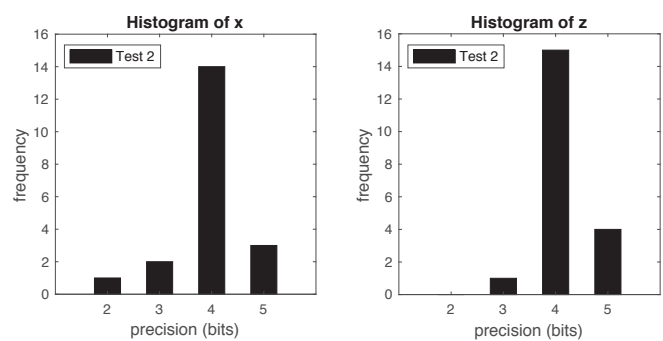

Fig. 11. Distributions of best fit precisions of wheels $x$ and $z$ for all participants from experiment 1 after passing nested hypothesis test 2.

If deviations from the mode are not statistically significant, a model in which the precision of the Prelec function is held constant could be a good fit. Such a model would have one less free parameter (allowing one extra degree of freedom). We note that a two-parameter continuous Prelec function is a special case of the three-parameter quantized Prelec function with the precision fixed at ∞. Therefore, by fixing the precision at 4 bits instead of allowing it to be a free parameter (i.e., $Q_4[.]Q_4[.]$ instead of $Q_k[.]Q_k[.]$), we eliminate one free parameter from the function. We repeated the nested hypothesis test with three additional 4-bit models included in our hypotheses (i.e., nested hypothesis test 2) and the results are shown in Table III. Overall, nested hypothesis test 2 produced two key findings: 70% of participants (i.e., 14/20) were non-homogeneous; 70% of participants (i.e., 14/20) were best fit to 4-bit models. The distributions of the best fit precisions of wheels $x$ and $z$ after nested hypothesis 2 are shown in Fig. 11, with modes of 4 bits.

### B. Interpretations of homogeneous and non-homogeneous participants

Typical probability distortion curves of a homogeneous and non-homogeneous participant are shown in Fig. 12. Mathematically, there is only one true answer to every conjunction probability. However, estimates of conjunctions are subjective, and vary among individuals. In our analysis, we used the quantized Prelec function to allow for these variations. Fig. 12A shows the probability distortion curve of a typical homogeneous participant in experiment 1. The dashed diagonal black line denotes the mathematical truth. Since both $x$ and $z$ are distorted similarly, there is only one distortion curve for each participant. Fig. 12B shows the distortion curves for a typical non-homogeneous participant, where $x$ and $z$ are distorted differently. Therefore, there is one distortion curve for $x$ and another for $z$; the curve for the "large" wheel (i.e., $x$) is plotted in red and the "small" wheel (i.e., $z$) in blue. Here, the same objective probability has more than one subjective truth within each individual.

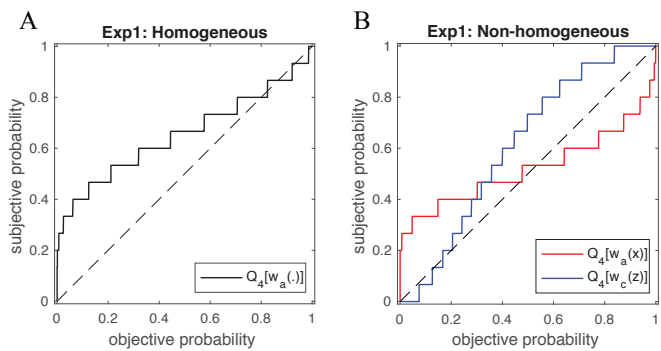

Fig. 12. The probability distortion curves of typical homogeneous and non-homogeneous participants of experiment 1. (A) $w_a$ represents the quantized Prelec function for the 2 wheels (i.e., $x$ and $z$); since this is a homogeneous participant, both wheels are distorted in the same manner, and therefore, there is only one common quantized Prelec function. (B) $w_a$ and $w_c$ represent the quantized Prelec functions for wheels $x$ and $z$ respectively.

The significance of subjective truth is further illustrated in Fig. 13. Consider a non-homogeneous participant from our experiment, presented with two separate cases of the 2-wheel task (Fig. 13A). In both cases, a wheel with 0.42 probability is presented, but its magnitude is different relative to the other wheel. This participant's probability distortion curves are shown in Fig. 13B. The vertical dotted line shows where 0.42 is objectively. This participant treats 0.42 differently in each case, as depicted by the two horizontal arrows in red and blue. In essence, not only does the truth/actuality lie subjectively in the mind of the beholder, there can be several shades of it internally.

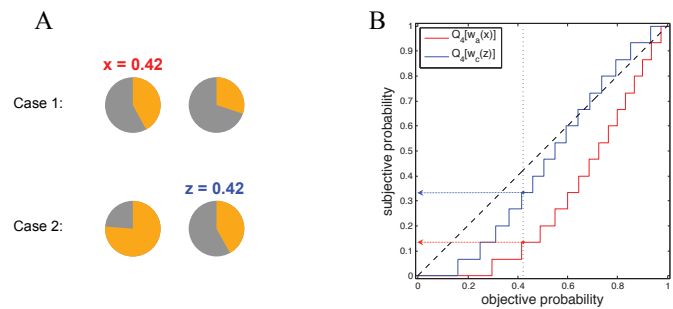

Fig. 13. An objective probability of 0.42 being treated differently depending on its relative magnitude to the other wheel. (A) Two separate cases of the 2-wheel task presented to the participant. (B) The corresponding distortion curves for the wheels $x$ and $z$, with the 2 arrows showing different subjective probabilities corresponding to the objective probability of 0.42.

### C. Experiment 2

Five out of 21 participants committed conjunction fallacy, which is consistent with experiment 1. The same data fitting and nested hypothesis tests were applied to the remaining 16 participants. Nested hypothesis tests results (detailed in Table IV) resemble those of experiment 1: 63% of participants (10/16) were non-homogeneous; 94% of participants (15/16) were best fit to 4-bit models. The distributions of best fit precisions of wheels $x$ and $z$ after nested hypothesis test 2 are shown in Fig. 14, with modes of 4 bits. These results (based on the 15 experimental conditions in Fig. 4), along with probability distortion curves (in Fig. 15), are consistent with those of experiment 1, indicating that the results of experiment 1 are generalizable and not specific to its 12 experimental conditions (in Fig. 3).

TABLE IV
RESULTS OF NESTED HYPOTHESIS TEST 2 FOR EXPERIMENT 2

| Models | Homogeneous? (i.e. same distortion function for x and z) | Same precision for x and z? | No. of free parameters | Degrees of freedom | No. of subjects passing nested test |
|---|---|---|---|---|---|
| $Q_4[x]Q_4[z]$ | Yes: linear | Yes: 4 bits | 0 | 6 | 0 |
| $Q_k[x]Q_k[z]$ | Yes: linear | Yes | 1 | 5 | 0 |
| $Q_k[x]Q_n[z]$ | Yes: linear | No | 2 | 4 | 0 |
| $Q_4[w_a(x)]Q_4[w_a(z)]$ | Yes: Prelec-2 | Yes: 4 bits | 2 | 4 | 5 |
| $Q_k[w_a(x)]Q_k[w_a(z)]$ | Yes: Prelec-2 | Yes | 3 | 3 | 0 |
| $Q_k[w_a(x)]Q_n[w_a(z)]$ | Yes: Prelec-2 | No | 4 | 2 | 1 |
| $Q_4[w_a(x)]Q_4[w_c(z)]$ | No: Prelec-2 | Yes: 4 bits | 4 | 2 | 10 |
| $Q_k[w_a(x)]Q_k[w_c(z)]$ | No: Prelec-2 | Yes | 5 | 1 | 0 |
| $Q_k[w_a(x)]Q_n[w_c(z)]$ | No: Prelec-2 | No | 6 | - | 0 |

There are 16 participants in total.

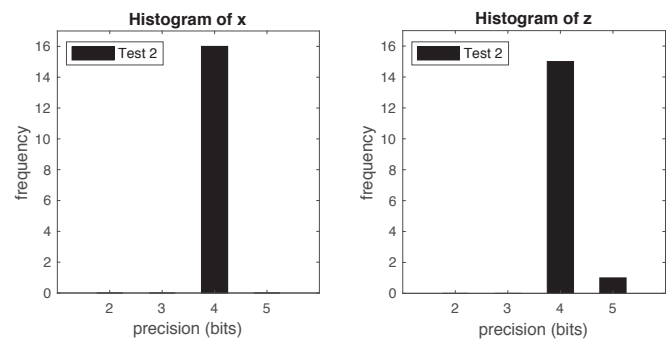

Fig. 14. Distributions of best fit precisions of wheels $x$ and $z$ for all participants from experiment 2 after passing nested hypothesis test 2.

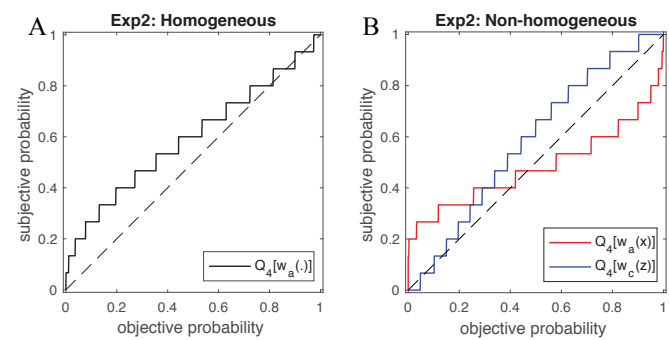

Fig. 15. The quantized probability distortion curves of typical homogeneous (A) and non-homogeneous (B) participants in experiment 2. These curves look consistent with those from experiment 1.

### D. Experiment 3

Four out of 44 participants committed conjunction fallacy, which is consistent with the findings of experiments 1 and 2. Similar data fitting and nested hypothesis tests were applied to the remaining 40 participants. Nested hypothesis tests results (described in Table V) resemble those of the 2-wheel experiments: 55% of participants (22/40) were non-homogeneous; 75% of participants (30/40) were best fit to 4-bit models. The distributions of best fit precisions of wheels $x$, $y$ and $z$ after nested hypothesis test 2 are shown in Fig.16, with modes of 4 bits. Typical probability distortion curves of a homogeneous and non-homogeneous participant are shown in Fig. 17, consistent with those from experiments 1 and 2.

TABLE V
RESULTS OF NESTED HYPOTHESIS TEST 2 FOR EXPERIMENT 3

| Models | Homogeneous? (i.e. same distortion function for x, y and z) | Same precision for x, y and z? | No. of free parameters | Degrees of freedom | No. of subjects passing the nested test |
|---|---|---|---|---|---|
| $Q_4[x]Q_4[y]Q_4[z]$ | Yes: linear | Yes: 4 bits | 0 | 9 | 1 |
| $Q_k[x]Q_k[y]Q_k[z]$ | Yes: linear | Yes | 1 | 8 | 0 |
| $Q_k[x]Q_m[y]Q_n[z]$ | Yes: linear | No | 3 | 6 | 0 |
| $Q_4[w_a(x)]Q_4[w_a(y)]Q_4[w_a(z)]$ | Yes: Prelec-2 | Yes: 4 bits | 2 | 7 | 10 |
| $Q_k[w_a(x)]Q_k[w_a(y)]Q_k[w_a(z)]$ | Yes: Prelec-2 | Yes | 3 | 6 | 3 |
| $Q_k[w_a(x)]Q_m[w_a(y)]Q_n[w_a(z)]$ | Yes: Prelec-2 | No | 5 | 4 | 4 |
| $Q_4[w_a(x)]Q_4[w_b(y)]Q_4[w_c(z)]$ | No: Prelec-2 | Yes: 4 bits | 6 | 3 | 19 |
| $Q_k[w_a(x)]Q_k[w_b(y)]Q_k[w_c(z)]$ | No: Prelec-2 | Yes | 7 | 2 | 2 |
| $Q_k[w_a(x)]Q_m[w_b(y)]Q_n[w_c(z)]$ | No: Prelec-2 | No | 9 | - | 1 failed all tests |

There are 40 participants in total.

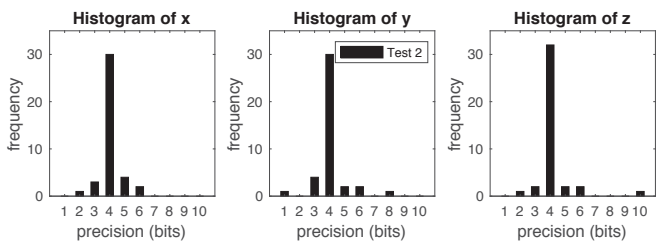

Fig. 16. Distributions of best fit precisions of wheels $x$, $y$ and $z$ for all participants from experiment 3 after passing nested hypothesis test 2.

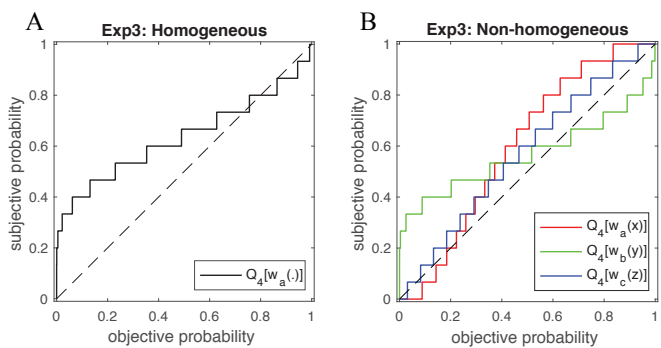

Fig. 17. The probability distortion curves of typical homogeneous and non-homogeneous participants of the 3-wheel experiment 3. (A) $w_a$ represents the quantized Prelec function for the 3 wheels (i.e., $x$, $y$ and $z$); since this is a homogeneous participant, all 3 wheels are distorted in the same manner. (B) $w_a$, $w_b$ and $w_c$ represent the quantized Prelec functions for wheels $x$, $y$ and $z$ respectively since this is a non-homogeneous participant.

### E. Summary of findings

Overall, 78% of all participants (i.e., 59/76) were best fit to 4-bit models (i.e., with 4 bits of quantization per wheel). The between-participant consistency across the 2- and 3-wheel tasks is quite remarkable, especially given that all participants in the experiments, bar one, are unique individuals. Fig. 18 shows a summary of the model parameters for all participants. There are no obvious clusters or patterns in the $\gamma$ and $\delta$ parameters values for all participants (Fig. 18A). The precision of all wheels across all participants range from 1 to 10 bits, with a mode of 4 bits (Fig. 18B). While the locations of the quantized steps (as determined by $\gamma$ and $\delta$) vary from participant to participant, the number of quantized steps in each wheel (as determined by $n$) is astonishingly similar (i.e., 4 bits = 16 steps), between participants, from wheel to wheel.

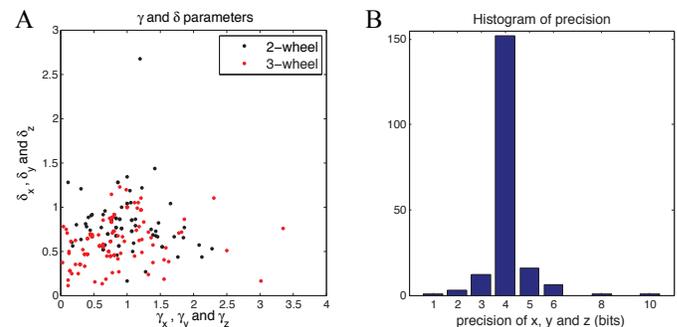

Fig. 18. Summary of the model parameters of all participants. (A) Scatterplot of $\delta$ and $\gamma$ parameters of the quantized probability distortion functions for all participants. (B) Histogram of the precision of all wheels across all participants. The precision of all wheels across all participants range from 1 to 10 bits, with a mode of 4 bits.

## V. DATA FITTING SAFEGUARDS

Our experimental setup and data analysis methodologies are complex in nature, such that biases in the methodologies could potentially lead us to arriving at the wrong conclusions. Therefore, it is prudent that we rigorously test our methodologies so as to mitigate these concerns. The key question is whether or not the experimental setup and the data fitting process are reliable and robust. That is, if a specific model (e.g., Model A) were to be used in making the choices/decisions in the experimental task, will the data fitting process result in the recovery of the same model (that made the choices/decisions in the experiment)? This model recovery question is illustrated in Fig. 19. If yes, then the experimental setup and data analysis methodologies are working as intended. With this goal in mind, three distinct safeguards have been employed as described in this section.

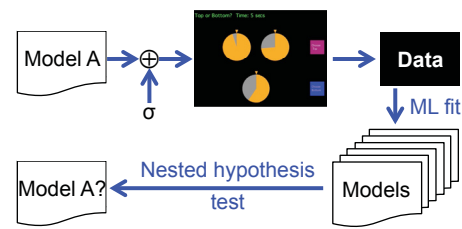

Fig. 19. Illustration of the model recovery question. If Model A was the decision maker in the experimental task, will it be recovered by the data fitting process?

## A. Checking the validity of data fitting process

To check whether or not the data fitting process is able to effectively detect different values of precision, we simulated participants having different precisions ranging from 2 to 20 bits. Here, we assume that 20 bits of precision represents the continuous model given that each step size is about $9.5 \times 10^{-7}$ (see Table I). Results of the simulation are summarized in Fig. 20. The x-axis shows the precision of the simulated participants (i.e., simulated decision makers) while the y-axis shows the corresponding precision obtained from the data fitting process. The data fitting process is able to correctly distinguish all simulated precisions up to 5 bits (as shown by the dashed red line), but it is unable to correctly distinguish simulated precisions of 6 bits and above without confounding, implying that the maximum detectable precision is limited to around 6 to 8 bits no matter how high the simulated precision is. This means that the data fitting process cannot tell the difference between, say, a 6-bit participant and a 16-bit participant. However, between 2 and 5 bits (which represents our range of interest), the fitting process is able to correctly detect the simulated precisions. Given that the majority of the participants in our experiment had 4 bits of precision, this simulation shows that the data fitting process was able to correctly detect our 4-bit findings.

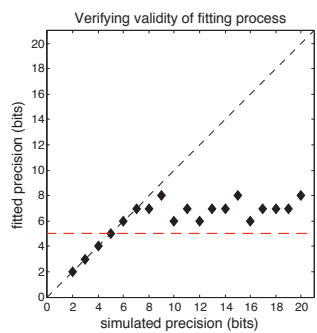

Fig. 20. Checking the validity of the fitting process. The x-axis shows the precision of simulated participants (i.e., simulated decision makers) while the y-axis shows the corresponding precision obtained from the fitting process. The dashed red line shows the limit of the simulation, above which precision becomes confounded.

## B. Confirming the reliability of data fitting process

To confirm that the fitting process is reliable (i.e., able to consistently detect what it is meant to detect), we performed bootstraps on two simulated participants. The first simulated participant had a 4-bit precision, whereas the second had a 20-bit precision. Each simulated participant was bootstrapped 100 times using the median value of the residual variances ($\sigma$) obtained from fitting actual experiment data. Results for the 4-bit simulated participant are shown in the top row of Fig. 21. In Fig. 21A and Fig. 21B, the vertical red lines show the $\gamma$ and $\delta$ parameter values of the simulated participant while the histogram shows the corresponding parameter estimates obtained from the 100 bootstrap runs. These histograms demonstrate that the data fitting process is able to consistently recover $\gamma$ and $\delta$ estimates that are close to the expected (i.e., simulated) values. The histogram of the precision estimates obtained from the bootstrap runs is plotted in Fig. 21C, showing that a 4 bit precision is consistently recovered in each of the 100 bootstrap runs. This gives us a false negative rate (i.e., Type II error rate) of less than 0.01 for a sample size of 100 bootstraps. The corresponding set of results for the 20-bit simulated participant are shown in the bottom row of Fig. 21. As in the previous case, the data fitting process is able to consistently recover $\gamma$ and $\delta$ estimates that are close to the expected values. However, when it comes to recovering the precision values, the data fitting process produced estimates ranging from 5 to 9 bits. This is expected given the results of the validity checks (Fig. 20). It also tells us that 5-bit precision estimates obtained from the data fitting process are slightly vulnerable to false positives (i.e., Type I errors, whereby a 20-bit simulated participant is mistakenly identified as a 5-bit participant). More importantly, none of the 100 bootstraps were mistakenly identified as 4 bits, which is where our main experimental results are concentrated. This gives us a false positive rate (i.e., Type I error rate) of less than 0.01 for the 4-bit precision estimates in our sample size of 100 bootstraps, indicating the reliability of our experimental data findings.

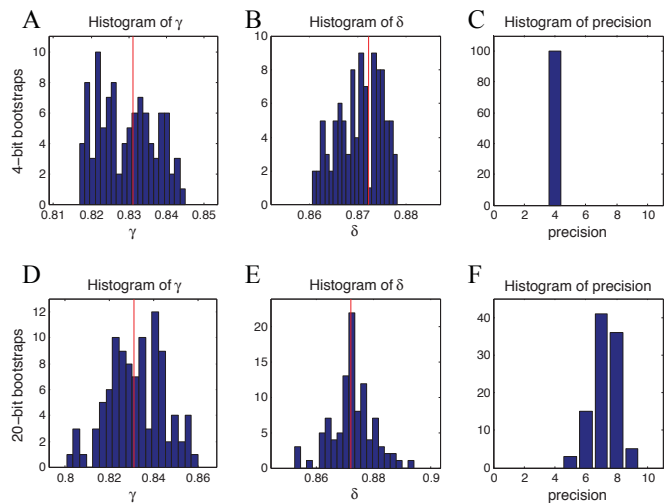

Fig. 21. Confirming the reliability of the data fitting process. (A, B, C) Histograms of fitted parameters for a simulated 4-bit homogeneous participant. (D, E, F) Histograms of fitted parameters for a simulated 20-bit homogeneous participant. In A, B, D and E, the vertical red lines show the corresponding expected (i.e., simulated) parameter values compared to the distribution of parameter estimates obtained from 100 bootstrap runs.

## C. Verifying the nested hypothesis testing process

To verify that the nested hypothesis tests are not confounding homogeneous and non-homogeneous participants, we simulated two homogeneous participants (i.e., 4-bit and 20-bit) and two non-homogeneous participants (4-bit and 20-bit). These simulated participants were passed through the fitting process and the nested hypothesis testing process to see whether the correct type (i.e., homogeneous or non-homogeneous) and precision are identified/recovered. The results of this simulation are shown in Table VI. All the four simulated participants were correctly identified in terms of type and precision (i.e., a simulated homogeneous participant is recovered by the homogeneous model, and a simulated non-homogeneous participant is recovered by the non-homogeneous model). Note that both 20-bit simulated participants were identified as having 7-bit precisions, which is expected given the conclusions from our previous reliability bootstraps (Fig. 21F). These results confirm that the nested hypothesis testing process performed as expected.

TABLE VI
VERIFYING THE NESTED HYPOTHESIS TESTING PROCESS

| Simulated subjects | | Fitted model | | | | Nested hypothesis test | | |
|---|---|---|---|---|---|---|---|---|
| | | Homogeneous (3 parameters) | | Non-homogenenous (5 parameters) | | | | Test result (best fit model) |
| Type | Simulated precision (bits) | Negative log likelihood ($\lambda_{H}$) | Fitted model's precision (bits) | Negative log likelihood ($\lambda_{NH}$) | Fitted model's precision (bits) | Test statistic: $2 * |\lambda_{NH} - \lambda_{H}|$ | $\chi^2$ threshold at 0.05 level | Reject simpler (homogeneous) model if test statistic > $\chi^2$ threshold |
| Homogeneous | 4 | -71.469 | 4 | -71.469 | 4 | 0 | 5.991 | Homogeneous 4-bit model |
| Homogeneous | 20 | -67.396 | 7 | -69.070 | 8 | 3.349 | 5.991 | Homogeneous 7-bit model |
| Non-homogeneous | 4 | -47.043 | 6 | -66.496 | 4 | 38.906 | 5.991 | Non-homogeneous 4-bit model |
| Non-homogeneous | 20 | -52.023 | 6 | -73.798 | 7 | 43.550 | 5.991 | Non-homogeneous 7-bit model |

Four simulated participants were passed through the data fitting process and the nested hypothesis testing process. The resultant best fit models matched the simulated participants, being classified as the same type and having the same expected precision.

Overall, these 3 safeguards demonstrate that our experimental setup and data analysis methodologies are valid, reliable and robust.

## VI. CONCLUSIONS

Our extensive work demonstrates that probability is represented in a discrete manner in the brain, and is very likely to be at 4 bits of precision, as opposed to the commonly assumed continuous representation. In order to accentuate the significance of our findings, Fig. 22 compares the best fit 4-bit quantized probability model (in black) with the best fit continuous probability model (in red) for a single sample participant. The experimental stimuli/conditions (i.e., wheels of the experiment) are plotted as dots. The gap between the black line and the red curve represents the effect size (i.e., the difference between the quantized model and the continuous model). For some experimental stimuli/conditions, the effect size is small (e.g., highlighted in blue), whereas for other experimental stimuli/conditions, the effect size is much larger (e.g., highlighted in green). Regardless of whether the effect sizes are small or large, the key point here is that a 4-bit quantized model makes a substantially different and more accurate prediction of human subjective probability compared to the continuous model.

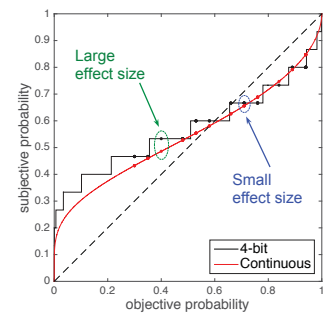

Fig. 22. Illustration of the effect sizes between the best fit 4-bit model and the best fit continuous model for one sample participant. The dots represent the experimental stimuli/conditions (i.e., wheels of the experiment).

A visual intuition for quantized probability is further illustrated in Fig. 23. In order to illustrate the intuition for quantization, we now introduce two terminologies: No Noticeable Difference (NND) and Big Noticeable Difference (BND). The first signifies each flat region of the quantized probability distortion curve (i.e., any value falling along this flat region is treated indifferently), whereas the second signifies the abrupt step between two flat regions of the curve (i.e., the value just after the step is significantly different from the value just before the step). Fig. 23A shows the quantized curve for a homogeneous participant from Experiment 3 (i.e., 3-wheel task). The experimental stimuli/conditions are plotted as dots. Recall that a homogeneous participant distorts all experimental stimuli/conditions (i.e., wheels $x$, $y$ and $z$) using the same quantized curve. The area highlighted by the dashed purple circle contains 3 dots. The second and third dots fall along the flat region, and therefore, both are treated indifferently from one another (i.e., No Noticeable Difference). The first dot falls on the flat region that is one step below, and therefore, it is treated differently to the second and third dots (i.e., Big Noticeable Difference). Fig. 23B shows the corresponding quantized curves and experimental stimuli/conditions for a non-homogeneous participant from Experiment 1.

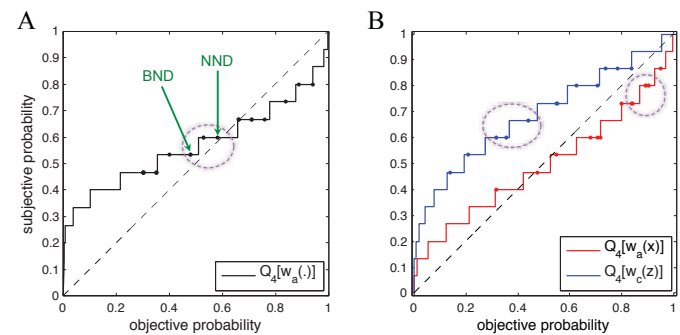

Fig. 23. Illustrations of No Noticeable Difference (NND) and Big Noticeable Difference (BND). The experimental stimuli/conditions (i.e., wheels of the experiment) are plotted as dots. (A) A homogeneous participant from Experiment 3 (i.e., 3-wheel task). (B) A non-homogeneous participant from Experiment 1 (i.e., 2-wheel task).

One example of the application of quantized probability lies in the interpretation of risks in decision-making. Suppose if the risk of an aircraft crashing due to engine failure were 10%, then, clearly, most people would choose not to fly on that aircraft. What if the risk was reduced to 5%? Or, 1%? Using a continuous representation of probability, one would conclude that people are more likely to fly if the probability of engine failure was 1%, compared to 5% or 10%. However, based on the current findings on quantized probability representation, we would conclude that people's decision on whether or not to fly will be the same (i.e., not more likely and not less likely) regardless of whether the probability was 1%, 5% or 10%. The decision would still (very) likely be "No", because people's subjective representation of probability for engine failure risk would be indifferent at the 1%, 5% and 10% probability range, and will likely remain so until the probability is perhaps reduced to the 0.001% range, or lower. In this sense, it is more important to find where the abrupt step change is. The location of the abrupt step change will also vary depending on the context/situation (e.g., the probability of getting lung cancer from smoking, the probability of a big snow storm hitting the city, the probability of a hurricane hitting and destroying a coastal town).

In our quantized model, each rise in the graph is an abrupt step rise. We suspect that, in reality, each rise may actually be an S-shaped rise, one that is consistent with the theories of psychometric functions and the Just Noticeable Difference (JND) threshold from Weber's law [25]. Certainly, the very notion of JND implies the existence of NND and BND – meaning, the JND is plausibly sandwiched somewhere in between the NND and the BND.

Our quantized probability representation conclusion here, derived from behavioral experimental data, supports and reinforces the theoretical findings of Tee and Taylor [17] which were derived from mathematical analyses and computer simulations – that, information in the brain is represented in a discrete form. As far as we know, our work here signifies the first and pioneering effort whereby a continuous model is compared and tested in an apples-to-apples equivalence fashion with its quantized counterpart. We end this paper with a generalized hypothesis that other forms of (cognitive) decisions made by the brain, such as intertemporal choice [26], are also likely to be quantized as well. Our generalized hypothesis can be tested in a rather straightforward manner by reanalyzing experimental data from existing and published results using the methodologies outlined here as the foundational blueprint.

ACKNOWLEDGMENT

J.T. thanks C. Randy Gallistel and Michael Woodford for their guidance, advice and feedback, which have been instrumental to the success of the work in this paper. J.T. acknowledges the contributions of Laurence T. Maloney on the acquisition of research funding for this work, and his inputs on the Data Fitting Safeguards in Section V. J.T. further thanks Laurence T. Maloney, Denis Pelli, and Jonathan Winawer, for their comments and suggestions on initial drafts of this work. J.T. and D.P.T. thank the late Bill Tranter for his constructive comments and suggestions.

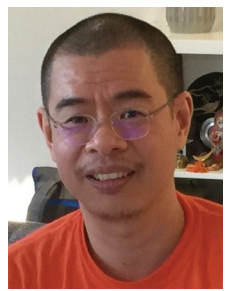

James Tee (M'17) completed his Ph.D. in Electrical & Electronic Engineering at the University of Canterbury in 2001, where he worked on Turbo Codes under the supervision of Des Taylor. Subsequently, he held various industry and policy positions at Vodafone Group, the World Economic Forum, New Zealand's Ministry of Agriculture & Forestry, and the United Nations. To facilitate his career transitions, he pursued numerous supplementary trainings, including an MBA at the Henley Business School, and an MPhil in Economics (Environmental) at the University of Waikato. In 2012, James began his transition into scientific research at New York University (NYU), during which he completed an MA in Psychology (Cognition & Perception) and a PhD in Experimental Psychology (Neuroeconomics). Most recently, he was an Adjunct Assistant Professor at NYU's Department of Psychology, and a Research Scientist (Cognitive Neuroscience) at Quantized Mind LLC. In September 2017, James began his venture into Mind-Body and Energy Medicine, where he is currently pursuing a 3-year MS in Acupuncture clinical training program to be an eastern medicine physician, at the Pacific College of Oriental Medicine in New York City. He is concurrently affiliated with the Communications Research Group at the University of Canterbury, where his research interests in neuroscience focuses on reverse engineering the communications codebook (i.e., signal constellation) of the Purkinje cell neuron. James is also pursuing research on Artificial Intelligence (AI) approaches inspired by insights drawn from psychology (cognition, perception, decision-making) and neuroscience.

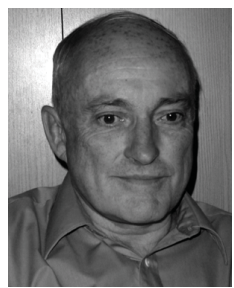

Desmond P. Taylor (LF'06) received the Ph.D. degree in electrical engineering from McMaster University, Hamilton, ON, Canada, in 1972. From 1972 to 1992, he was with the Communications Research Laboratory and the Department of Electrical Engineering, McMaster University. In 1992, he joined the University of Canterbury, Christchurch, New Zealand, as the Tait Professor of communications. He has authored approximately 250 published papers and holds several patents in spread spectrum and ultra-wideband radio systems. His research is centered on digital wireless communications systems focused on robust, bandwidth-efficient modulation and coding techniques, and the development of iterative algorithms for joint equalization and decoding on fading, and dispersive channels. Secondary interests include problems in synchronization, multiple access, and networking. He is a Fellow of the Royal Society of New Zealand, the Engineering Institute of Canada, and the Institute of Professional Engineers of New Zealand.